# The DCT Model as a Novel Regression Framework within a Lagrangian Formulation

Marc Martinez-Gost[1], Ana I. Perez Neira[1,2] , Miguel Ángel Lagunas[2]
[1] Centre Tecnològic de Telecomunicacions de Catalunya, Castelldefels, Spain
[2] Universitat Politècnica de Catalunya, Barcelona, Spain
{mmartinez, aperez, m.a.lagunas}@cttc.es

*Abstract*—**This paper introduces a unified regression framework based on the Lagrange formalism, demonstrating how polynomial and logistic regression can all be formulated within a common variational (Lagrangian formalism) structure. Within this framework, the DCT-based (Discrete Cosine Transform) model naturally emerges as a novel and effective approach to traditional or unsupervised regression. The DCT is used as the constraints in the Lagrangian formalism. By leveraging the nearly orthogonal and bounded nature of the cosine basis, the DCT model offers computational advantages and improved convergence properties compared with traditional polynomial methods. The results further support the potential of the DCT-based neuron as a powerful tool for regression analysis and related learning tasks.**

## I. INTRODUCTION

Regression analysis is probably the topic that, over time, has received one of the largest lists of published papers and books. For this reason, describing all the previous work in the field will be almost impossible for a reduced length manuscript. Even selecting the most relevant contributions is also difficult since there are excellent books where all these relevant contributions are included. To the authors knowledge, references [1]-[3] describe all these preliminary works performed in the field.

The major contributions of this paper are twofold. First, we introduce a unified variational (Lagrangian) formulation of regression problems, in which different regression models are obtained by minimizing an objective function subject to data-driven constraints. Within this framework, classical approaches such as polynomial (e.g., Volterra-type), and logistic models can be interpreted as particular cases arising from specific choices of objective functions and constraint sets (i.e., moment-based or spectral constraints). Second, we extend this formulation by incorporating constraints expressed in the Discrete Cosine Transform (DCT) domain, providing an alternative spectral representation of the data that leads to a different class of regression models with improved numerical properties. This work builds upon the DCT-based perceptron introduced in [4] and [5], where bounded and orthogonal cosine-based kernels lead to faster convergence in gradient-based learning for supervised function regression and classification tasks. In the present paper, these ideas are generalized to a variational regression framework that unifies classical and spectral approaches and highlights how different combinations of objectives and constraints naturally generate new regression formulations.

The paper structure starts with a brief and well-known description of the linear regression in Section II. Section III contains the same as the previous section but for the case of logistic regression, including the computation of the parameters with an iterative method based on the statistical gradient. The choice of the algorithm is based on its computational complexity and not on its performance for polynomial logistic regression. There are methods based on Lagrange formulations [6], and specifically using the Fourier transform with a different objective that depends on the derivatives of the function [7]. We will assume that the constraints will be linear in terms of the regression function $f(\cdot)$. In addition, the objective will not include derivatives of function $f(\cdot)$. These two assumptions reduce the scope of the procedure but greatly simplify the complexity of the problem and ensure the uniqueness of the solution.

Without loss of generality, the linear and polynomial regression in this paper focusses on the case of a single explanatory variable $x$ and a single dependent variable $y$. For logistic regression, it considers a single explanatory variable for the case of only two categories, without loss of generality as well. Section IV introduces the DCT model by just changing the set of constraints and develops the DCT model for linear and polynomial regression. Section V describes the use of the DCT model for the logistic regression. Conclusions and references end the document.

The Lagrangian formulation refers hereafter to the use of a function $\psi(\cdot)$ for the objective together with a set of, let us say, $M$ constraints using functions (kernels) $\phi_m(\cdot)$, for $m = 1, \dots, M$. This is formulated in (1). The $M$ constraints come from the $N$ data available $(x_1, \dots, x_N)$ and the corresponding dependent variables $(y_1, \dots, y_N)$. For this reason, in terms of Machine Learning or Pattern recognition, regression can be considered as a supervised procedure. Reference [8] is a nice review of the state of the art in regression and the links with Machine Learning.

$$\text{minimize} \quad \int \psi\big(f(x)\big)dx \tag{1a}$$

$$\text{subject to} \quad \int \phi_m\big(f(x)\big)dx = \beta_m\,, \qquad m = 1, \dots, M \tag{1b}$$

Note that function selected for the objective is merely a "cosmetic" choice. For example, selecting $\psi\big(f(x)\big) =$

$f(x)^2$ is equivalent to choosing the minimization of the energy of the function and, alternatively, selecting $\psi(f(x)) = \ln f(x)$ is equivalent to looking for the maximum flatness function, whenever the average of the function $f(x)$ is constrained (i.e., for some $\phi_m(f(x)) = 1$). In summary, the function of the objective indicates which feature (cosmetic) we prefer to stand out from the infinite solutions of possible functions that hold the constraints. We are assuming that the number of constraints $M$ is always less than the number of input-output pairs, $N$. The case of $M = N$ is denoted as the saturated model.

Problem (1) is solved by first forming the Lagrangian as shown in (2a) and then differentiating it with respect to $f(x)$, as in (2b). The function $f(x)$ is obtained by setting (2b) equal to zero, while the Lagrangian multipliers $\lambda_m$ are determined from the constraints. The constraint values $\beta_m$ are derived from the dataset.

$$\mathcal{L}(f(x), \lambda_1, \dots, \lambda_M) = \int \psi(f(x))dx - \sum_{m=1}^{M} \lambda_m \left( \int \phi_m(f(x))dx - \beta_m \right) \quad (2a)$$

$$\frac{\partial \mathcal{L}}{\partial f(x)} = \frac{\partial \psi(f(x))}{\partial f(x)} - \sum_{m=1}^{M} \lambda_m \phi_m(f(x)) \quad (2b)$$

The DCT model was introduced for feedforward neural networks with adaptive activation functions [4]. The DCT model of a nonlinearity $g(x)$ is modeled as (3a), where the $c_q$ are the DCT coefficients of $g(x)$ and $N_{DCT}$ is the length of the sampled $g(x)$. The DCT model or DCT regression means that the constraints are formulated with functions as (3b) as it will be described hereafter. Its role in regression will be included in sections IV and V for polynomial and logistic regression respectively.

$$g(x) \approx \sum_{q=1}^{Q} c_q \cos\left( \frac{\pi}{2N_{DCT}} (q-1)(2x-1) \right) \quad (3a)$$

$$\phi_m(x) = \cos\left( \frac{\pi}{2N_{DCT}} (q-1)(2x-1) \right) \quad (3b)$$

As mentioned, with DCT in regression, polynomial or logistic, it refers just to use cosine terms in the constraints. Nevertheless, the success of these kernels in regression roots in the performance of low complexity and full control of the tradeoff between convergence rate and miss-adjustment on iterative methods [4].

Finally, we remark that the continuous variational formulation is included for conceptual clarity, while all practical developments are carried out in the discrete setting. In this case, integrals are replaced by finite sums over the available samples $x_n$, $n = 1, \dots, N$. This discretization does not alter the essence of the variational problems in (2), provided that the objective does not involve derivatives and that the constraints are linear equality conditions in the target function.

## II. LINEAR REGRESSION

Given $N$ data of explanatory variable $(x_1, \dots, x_N)$ and the corresponding dependent variable values $(y_1, \dots, y_N)$, among the infinite functions that hold the $M$ $(M < N)$ constraints, we select the minimum energy solution as the objective. The constraints will be the first and second moments of $f(\cdot)$ using the dependent variables data, i.e., $\phi_1(x) = 1$ and $\phi_2(x) = x$. The overall optimization problem is:

$$\text{minimize} \quad \sum_{n=1}^{N} f(x_n)^2 \quad (4a)$$

$$\text{subject to} \quad \sum_{n=1}^{N} f(x_n) = \sum_{n=1}^{N} y_n \quad (4b)$$

$$\sum_{n=1}^{N} x_n f(x_n) = \sum_{n=1}^{N} x_n y_n \quad (4c)$$

Following the procedure in (2), results in a linear dependence of $f(\cdot)$ with respect to the explanatory variable:

$$f(x_n) = \lambda_1 + \lambda_2 x_n, \quad n = 1, \dots, N \quad (5)$$

In order to find the Lagrange multipliers from the constraints, we compute (6):

$$\begin{bmatrix} N & \sum_{n=1}^{N} x_n \\ \sum_{n=1}^{N} x_n & \sum_{n=1}^{N} x_n^2 \end{bmatrix} \begin{bmatrix} \lambda_1 \\ \lambda_2 \end{bmatrix} = \begin{bmatrix} \sum_{n=1}^{N} y_n \\ \sum_{n=1}^{N} x_n y_N \end{bmatrix} \quad (6)$$

In general, for an order $M$, the model for $f(x)$ will be (7a) and the corresponding normal equations are (7b). When extending the formulation from two constraints to the general case of $M$ constraints, the coefficient matrix correspondingly changes from a $2 \times 2$ matrix to an $M \times M$ matrix.

$$f(x) = \sum_{m=0}^{M-1} \lambda_m x^m \quad (7a)$$

$$\begin{bmatrix} N & \sum_{n=1}^{N} x_n & \cdots & \sum_{n=1}^{N} x_n^{M-1} \\ \vdots & \vdots & & \vdots \\ \sum_{n=1}^{N} x_n^{M-1} & \sum_{n=1}^{N} x_n^{M} & \cdots & \sum_{n=1}^{N} x_n^{2(M-1)} \end{bmatrix} \begin{bmatrix} \lambda_1 \\ \lambda_2 \\ \vdots \\ \lambda_{M-1} \end{bmatrix} = \begin{bmatrix} \sum_{n=1}^{N} y_n \\ \sum_{n=1}^{N} x_n y_N \\ \vdots \\ \sum_{n=1}^{N} x_n^{M-1} y_N \end{bmatrix} \quad (7b)$$

Note twofold, first the form of $f(x)$ is settled by the constraints. Second, the obtained result is the same as the direct minimization of the mean square error (MSE) for linear regression. In fact, the MSE when the mean is constrained coincides with the minimum energy objective. Polynomial regression results when the number of constraints $M$ is higher than 2, i.e., including other moments using $\phi_m(x) = x^m, m = 0,..,M-1$.

We will use a set of data commonly used in any basic course on regression, shown in Table 1. The explanatory variable will be the number of hours devoted by a student and the dependent variable will be the grade (0-10) of the evaluation. The length of the data is 9. The resulting linear regression is shown in Figure 1.

**Table 1.** Dataset for polynomial regression.

| $x_n$ | 1.0 | 1.7 | 2.5 | 3.4 | 3.8 | 4.1 | 5.0 | 5.5 | 6.2 |
|---|---|---|---|---|---|---|---|---|---|
| $y_n$ | 1.2 | 2.8 | 1.2 | 3.6 | 5.3 | 3.5 | 3.2 | 4.7 | 6.5 |

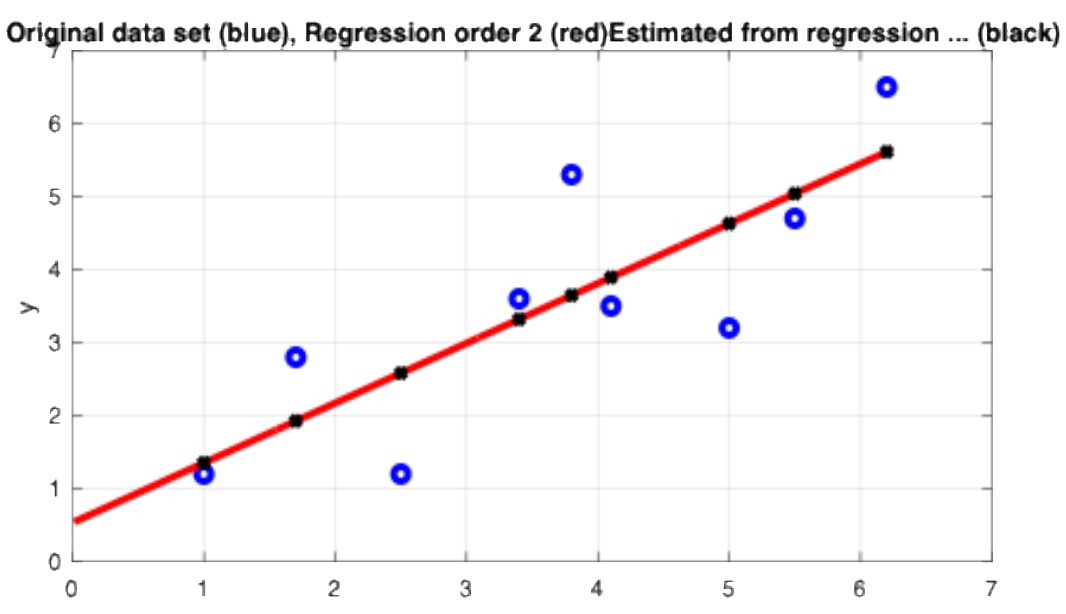

**Figure 1.** Linear regression (red), data (blue circles), predicted (black asterisk).

A brief review of quality in linear regression follows. There are many methods, and among them, hereafter a specific set of parameters will be selected just to compare the different methods. Currently there is not a clear consensus about the so called $p$-value. For this reason, we will only include the factors $R^2$ and $F$ with an intuitive explanation for its use.

To check the value of the obtained regression we compare the MSE between the dependent data $y_n$ and the value provided by the regression curve $f(x_n)$, termed "residuals". The sum of the squared residuals provides the value for $\mathrm{MSE_{fit}}$. Next, we compare with the MSE when there is not explanatory variable. In this case the regression will be a constant equal to the mean of the dependent variable $y_n$, named $\mathrm{MSE_{mean}}$. Both metrics are defined below:

$$\mathrm{MSE_{fit}} = \frac{1}{N}\sum_{n=1}^{N}(f(x_n) - y_n)^2 \qquad (8a)$$

$$\mathrm{MSE_{mean}} = \frac{1}{N}\sum_{n=1}^{N} y_n \qquad (8b)$$

Finally, we also define $\mathrm{MSE_{sat}}$, as the difference between the mean and the saturated model. Both in the polynomial and in the DCT model, included hereafter, this $\mathrm{MSE_{sat}}$ is zero and it will not be included in our formulation. With these values, the so-called $R^2$ factor is obtained as (9a). Clearly $R^2$ explains how valuable it is to include the explanatory variable in the regression. $R^2$ close to 1 (100%) means that the explanatory variable $x_n$ describes perfectly the dependent variable $y_n$. Irrelevant variables will produce almost zero value (0%) on $R^2$. This test is quite useful to decide to include additional explainable variables or not. From the example of Figure 1, $R^2 = 0.65$. Clearly $R^2$ will decrease whenever $M$ increases, and we have $R^2 = 1$ for $M = N$. In general, we want the $\mathrm{MSE_{fit}}$ to be larger than $\mathrm{MSE_{mean}}$ but not too close to the saturated model. In consequence, it is necessary to introduce a parameter that reflects how the increase in degrees of freedom (DoF) $M$ reduces the quality. This is the $F$-factor, like the $R^2$ but divided by the difference of DoF, as shown in (9b). Note that $F = 0$ for $M = N$.

$$R^2 = \frac{\mathrm{MSE_{mean}} - \mathrm{MSE_{fit}}}{\mathrm{MSE_{mean}} - \mathrm{MSE_{sat}}} = \frac{2.758 - 0.956}{2.758} = 0.65 \quad (9a)$$

$$\mathrm{F} = \frac{\dfrac{\mathrm{MSE_{mean}} - \mathrm{MSE_{fit}}}{\mathrm{DoF_{fit}} - \mathrm{DoF_{mean}}}}{\dfrac{\mathrm{MSE_{mean}} - \mathrm{MSE_{sat}}}{\mathrm{DoF_{sat}} - \mathrm{DoF_{fit}}}} = \frac{\dfrac{2.758 - 0.956}{2-1}}{\dfrac{0.956}{9-2}} = 4.57 \quad (9b)$$

For polynomial regression, we analyze the cases of $M = \{2, 3, 4, 5\}$. The indexes of quality are shown in Table 2 and some results depicted in Figure 2. Notice that for $M = 0$, we have $\mathrm{MSE_{fit}} = \mathrm{MSE_{mean}}$ and $\mathrm{MSE_{sat}} = 0$.

**Table 2.** Results for polynomial regression.

| $M$ | $\mathrm{MSE_{fit}}$ | $R^2$ | $F$ | rcond |
|---|---|---|---|---|
| 1 | 6.06 | - | - | - |
| 2 | 0.76 | 0.87 | 8.70 | $5.1\cdot10^{-3}$ |
| 3 | 0.73 | 0.87 | 3.95 | $2.2\cdot10^{-5}$ |
| 4 | 0.72 | 0.88 | 2.34 | $7.8\cdot10^{-8}$ |
| 5 | 0.72 | 0.88 | 1.53 | $2.0\cdot10^{-10}$ |

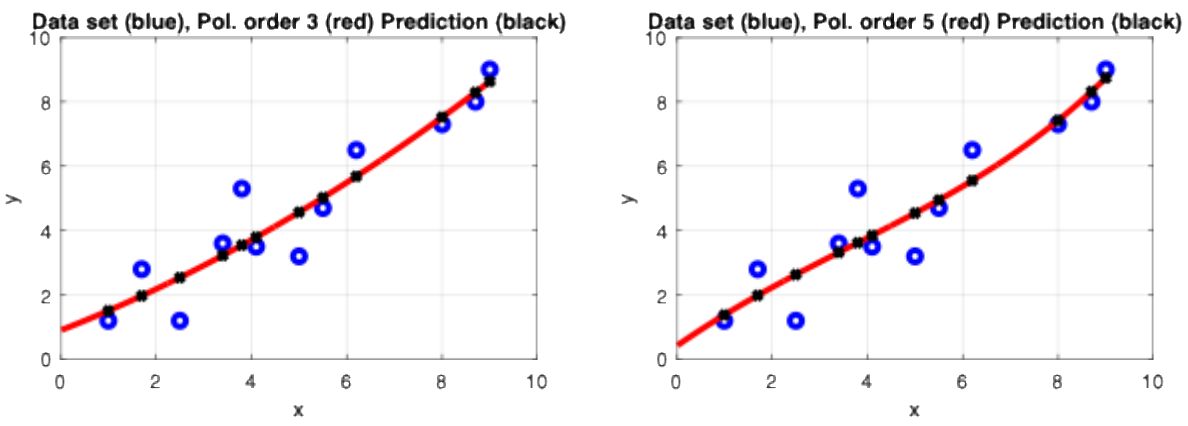

**Figure 2.** Polynomial regression (left) $M = 2$ and (right) $M = 5$.

Clearly a linear regression is enough to represent the curve $f(x)$ in this example. This is motivated by the fact that additional coefficients besides $\lambda_1$ are close to zero. The degradation of $F$ means that the statistical value of increasing the number of variables decreases. It is important to remark that the reciprocal condition number (rcond) of the matrix is extremely low. This implies an extremely high sensitivity to small changes (e.g., noise) in the values of the dependent variable. Furthermore, for values less than 1 the prediction is highly negative precluding any prediction outside interval covered by the explanatory variable. Effects like robustness to noisy dependent variable are a syndrome of the up to what degree parameter $R^2$ and $F$ are adequate to measure the statistical value of the model.

## III. POLYNOMIAL LOGISTIC REGRESSION

The difference with linear regression is that in this case the dependent variable $y$ is discrete, taking values that represent the categories. In the case of two categories this will be designated as "0" and "1". The selected dataset for logistic regression is shown in Table 3, also used in basic courses on logistic regression.

**Table 3.** Dataset for logistic regression.

| $x_n$ | 0.5 | 0.7 | 1 | 1.25 | 1.5 | 1.75 | 2 | 2.25 | 2.5 | 2.75 |
|---|---|---|---|---|---|---|---|---|---|---|
| $y_n$ | 0 | 0 | 0 | 0 | 0 | 0 | 0 | 1 | 0 | 1 |
| $x_n$ | 3 | 3.25 | 4 | 4.25 | 4.5 | 4.75 | 5 | 5.5 | 5.75 | |
| $y_n$ | 0 | 1 | 0 | 1 | 1 | 1 | 1 | 1 | 1 | |

The objective is to find a function $p(x)$ that provides the probability that $x$ belongs to category "1" and $1 - p(x)$ representing the probability for category "0". At a given $x$, the entropy of the distributions is

$$H\big(p(x)\big) = -p(x)\ln p(x) - \big(1 - p(x)\big)\ln\big(1 - p(x)\big) \quad (10)$$

To obtain $p(x)$, the global objective is that the sum of all the entropies at the explanatory variable will be maximum or, without the negative sign, minimum as in (11). The maximum-entropy principle justifies choosing $p(x)$, as the distribution with maximum entropy is the less biased one, introducing no structure beyond what the constraints required. The constraints will be the moments, as previously, since $f(x)$ represents the probability function. The overall optimization problem is:

$$\text{minimize} \quad -\sum_{n=1}^{N} H\big(p(x_n)\big) \quad (11a)$$

$$\text{subject to} \quad \sum_{n=1}^{N} x_n^m p(x_n) = \sum_{n=1}^{N} x_n^m y_n \, , m = 0, \dots, M-1 \quad (11b)$$

From the Lagrangian, the partial derivatives obtained from (11) are

$$\frac{\partial \mathcal{L}}{\partial p(x_n)} = \ln p(x_n) + 1 - \ln\big(1 - p(x)\big) - 1 - \sum_{m=1}^{M} \lambda_m x_n^m \quad (12)$$

Notice that the second derivative, shown in (13), reveals that the optimum distribution corresponds to a minimum, given that $0 \leq p(x_n) \leq 1$.

$$\frac{\partial^2 \mathcal{L}}{\partial p(x_n)^2} = \frac{1}{p(x_n)\big(1 - p(x_n)\big)} > 0 \quad (13)$$

Forcing the derivative in (12) equal to zero we obtain (14a). This formula reveals that the logistic regression coincides with fitting a linear model of the quantity $\ln(p/(1-p))$ (i.e., the logarithm of "odds" or logits). This can be viewed as the log-likelihood (LL) ratio of probabilities $p(y = 1|x)$ and $p(y = 0|x)$. Rearranging (14a) results in (14b), which provides the final expression for $p(x_n)$.

$$\ln \frac{p(x_n)}{1 - p(x_n)} = \sum_{m=0}^{M-1} \lambda_m x_n^m \quad (14a)$$

$$p(x_n) = \frac{1}{1 + \exp(-\sum_m \lambda_m x_n^m)} \quad (14b)$$

Note that (14b) can be viewed as a perceptron, used for classification, where the input are polynomial kernels, and the nonlinearity is a sigmoid. Nevertheless, the logistic is not by itself a classifier, but can be used for this purpose by just setting a threshold to the probability, i.e. for the threshold set to 0.5, for $p(x) = 0.7$, the classifier will provide that the dependent variable belongs to category "1" with confidence equal to 70%.

The results for $M = 2$ and $M = 5$ are shown in Figure 3. Note that when the order increases, the regression curve evidences and increases attention to the outliers. The computation of the coefficients in (12), as well as the parameters of quality will be presented hereafter in subsection III-1.

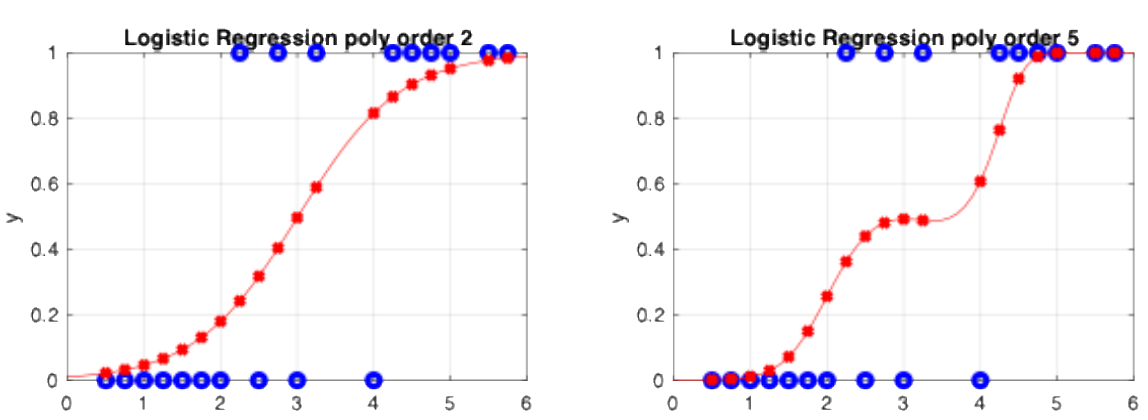

**Figure 3.** Logistic polynomial regression (left) $M = 2$ and (right) $M = 5$.

### A. *Computing Coefficients and Quality Parameters*

Note that enforcing the constraints in (11) does not yield a closed-form solution for the Lagrangian operators. For this reason, instead of imposing the constraints, we will design them such that the so-called log-loss (15a) is minimized. This is equivalent to minimizing the cross-entropy in (15b) between the dependent variable and the logistic regression, where the numerical values of the categories are considered as a probability, i.e., $p(y_n = 1) = 1$ and $p(y_n = 0) = 0$. The cross-entropy formulation provides a closed-form gradient expression, which facilitates learning the Lagrangian operators from data through a stochastic gradient algorithm. In addition, (15b) could be viewed as the negative LL of the random variable $x$, which follows a Bernoulli distribution. In other words, minimizing the cross-entropy will be equivalent to maximizing the LL.

$$\mathcal{L}_{log} = -\sum_{n=1, y_n=1}^{N} \ln p(x_n) - \sum_{n=1, y_n=0}^{N} \big(1 - p(x_n)\big) \quad (15a)$$

$$\mathcal{L}_{ce} = -\sum_{n=1}^{N} y_n \ln p(x_n) + (1-y_n)\ln\big(1-p(x_n)\big) \quad (15b)$$

The gradient of the cross-entropy in (15b) is (16a), where we express $\partial p(x_n)/\partial\lambda_m = p(x_n)\big(1-p(x_n)\big)$. The stochastic gradient algorithm, presented in (16b), updates the coefficients sequentially at every $n = 1, \dots, N$.

$$\frac{\partial \mathcal{L}_{ce}}{\partial \lambda_m} = -\sum_{n=1}^{N} x_n^m \big(y_n - p(x_n)\big) \quad (16a)$$

$$\lambda_m \longleftarrow \lambda_m - \mu \frac{\partial \mathcal{L}_{ce}}{\partial \lambda_m} = \lambda_m + \mu\left(x_n^m\big(y_n - p(x_n)\big)\right) \quad (16b)$$

Note that in order to update the coefficients with the stochastic gradient, the polynomials make the dynamic range of the gradients very large and the convergence uncertain. It is possible to select the step size equal to $\mu = \alpha(M)/M$, where $\alpha$ depends on $M$. The unbounded character of the kernels and their correlation forces to a fine tuning of $\alpha$ for each order $M$ to get a good trade-off between convergence rate, i.e., number of iterations, and miss-adjustment.

Regarding the quality parameters, we take the LL ratio ($\mathrm{LL_{fit}}$), corresponding to (15b), to replace the MSE of the polynomial regression. In consequence the resulting $R^2$ is the same as in linear regression, but it is expressed in terms of LL (McFaddens' pseudo $R^2$). The $\mathrm{LL_{mean}}$ is now termed $\mathrm{LL_{op}}$ and denotes the overall probability. All LL are negative values and $\mathrm{LL_{fit}}$ will be greater or equal to $\mathrm{LL_{op}}$. These definitions are analogous to (8). Accordingly, the $F$-factor used for logistic regression will be (17b).

$$R^2 = \frac{\mathrm{LL_{fit}} - \mathrm{LL_{op}}}{\mathrm{LL_{sat}} - \mathrm{LL_{op}}} \quad (17a)$$

$$\mathrm{F} = \frac{\dfrac{\mathrm{LL_{fit}} - \mathrm{LL_{op}}}{M-1}}{\dfrac{\mathrm{LL_{sat}} - \mathrm{LL_{op}}}{N-M}} \quad (17b)$$

Table 4 shows the configuration and results for logistic polynomial regression for different model orders. Notice that $\mathrm{LL_{fit}} = \mathrm{LL_{op}}$ for $M = 1$.

**Table 4.** Configuration and convergence for polynomial regression.

| $M$ | $\alpha$ | Iterations | $\mathrm{LL_{fit}}$ | $R^2$ | $F$ |
|---|---|---|---|---|---|
| 1 | - | - | -13.17 | - | - |
| 2 | $10^{-2}$ | $3\cdot10^{3}$ | -6.62 | 0.49 | 8.93 |
| 3 | $10^{-2}$ | $1.5\cdot10^{5}$ | -6.58 | 0.49 | 4.49 |
| 4 | $10^{-3}$ | $2\cdot10^{6}$ | -5.82 | 0.55 | 3.34 |
| 5 | $10^{-4}$ | $2\cdot10^{7}$ | -4.89 | 0.62 | 2.28 |

Clearly the stochastic gradient algorithm does not work properly with the problem of minimizing the cross-entropy for the polynomial regression. The convergence rate decreases severely when the order increases. As mentioned before, this is motivated by the extreme dynamic range of the kernels and the kernels are correlated.

## IV. DCT MODEL IN REGRESSION

The main advantage of introducing the Lagrange formalism is that both the cosmetic objective and the constraints can be modified without changing the overall procedure. The DCT model introduced in (3) arises from preserving the same objective, while adopting a new set of constraints. Particularly, the new set of constraints is obtained substituting the polynomial model for the DCT of $f(\cdot)$ at $M$ frequency components. The corresponding optimization problem is shown in (17).

$$\text{minimize} \sum_{n=1}^{N} f(x_n)^2 \quad (18a)$$

$$\text{subject to} \sum_{n=1}^{N} f(x_n)\cos\left(\frac{\pi(m-1)}{2N_{DCT}}(2z_n - 1)\right) = \sum_{n=1}^{N} y_n \cos\left(\frac{\pi(m-1)}{2N_{DCT}}(2z_n - 1)\right), m = 1, \dots, M \quad (18b)$$

In (18b) variable $z_n = \left(\frac{N_{DCT}-1}{x_{max}}\right)x_n$ represents a linear mapping from the domain of $x_n$, namely $0 \le x_n \le x_{max}$, to the domain required by the DCT, which is $0 \le z_n \le N-1$. Furthermore, notice that $m$ starts at 1, which corresponds to the first moment. Besides this term, the rest do not correspond to moments.

Note that when $f(x_n)$ is an even function, i.e., $f(x_n) = f(-x_n)$, the constraints become like fixing the DCT of $f(\cdot)$ at its first $M$ values. The evenness character is not a restrictive, and DCT provides a more compact set of significant values around the origin than the discrete Fourier transform (DFT).

The resulting DCT model from (18) is (19).

$$f(x_n) = \sum_{m=1}^{M} \lambda_m \cos\left(\frac{\pi(m-1)}{2N_{DCT}}\left(2\left(\frac{N_{DCT}-1}{x_{max}}\right)z_n - 1\right)\right) \quad (19)$$

One of the most relevant features of the DCT is that its kernels are orthogonal and bounded. These properties imply that gradient based algorithm leads to simpler designs than in the polynomial case. This is due to diagonal structure of the matrix in the normal equations, with entries equal to 1/2, this is, the power of the cosine kernels. Due to this uncorrelation between coefficients, when moving from a model of order $M$ to one of order $M+1$, the first $M$ coefficients remain unaltered. This means that, even high order models, the LMS algorithm does not need to be re-tuned to order $M$, unlike in polynomial regression.

The performance for the DCT regression is shown in Table 5 and the resulting DCT regressions for orders $M = 2$ and $M = 5$ are shown in Figure 4. The performance of the DCT in regression is similar to the polynomial model. The difference is small and, in both cases, it is evident that the data are well described by a model order 2. For predictions outside the data interval, the bounded nature of kernels in DCT provides better results. In summary the difference is basically cosmetic. At the same time the condition numbers for the DCT model are more reasonable than those for the case of using polynomial regression.

**Table 5.** Results for DCT regression.

| $M$ | $\text{MSE}_{\text{fit}}$ | $R^2$ | $F$ | rcond |
|---|---|---|---|---|
| 1 | 6.06 | - | - | - |
| 2 | 0.95 | 0.84 | 8.39 | 0.1 |
| 3 | 0.86 | 0.86 | 3.86 | 0.2 |
| 4 | 0.75 | 0.87 | 2.33 | 0.26 |
| 5 | 0.75 | 0.87 | 1.54 | 0.39 |

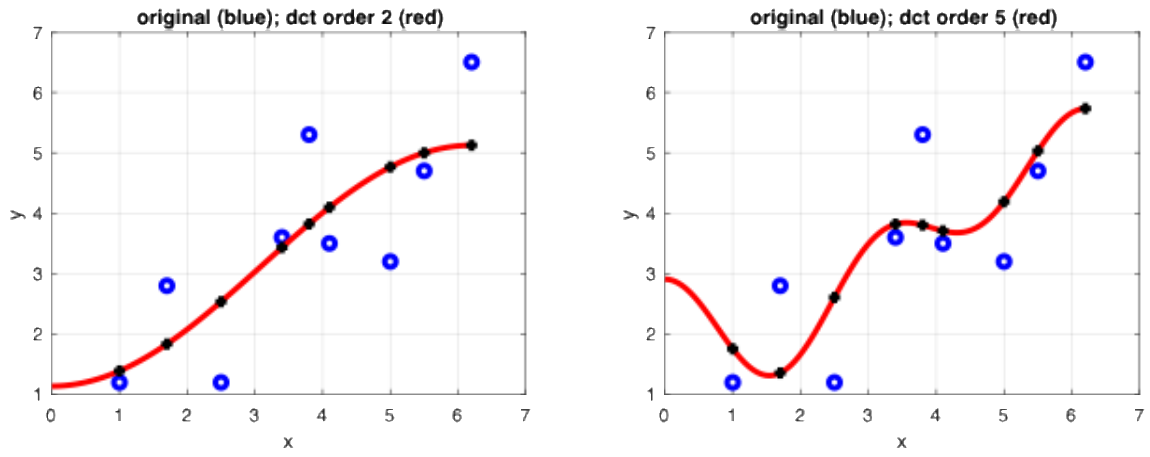

**Figure 4.** DCT regression. (Left) $M = 2$ and (Right) $M = 5$.

*A. DCT in Logistic Regression*

Next, we will analyze the case of logistic regression under polynomial and DCT models. As we did for standard regression, we can systematically take the polynomial logistic regression problem in (11) and substitute the DCT model. This results in (20).

$$\text{minimize} \quad -\sum_{n=1}^{N} H\big(p(x_n)\big) \tag{20a}$$

$$\text{subject to} \sum_{n=1}^{N} f(x_n)\cos\left(\frac{\pi(m-1)}{2N_{DCT}}(2z_n-1)\right) = \sum_{n=1}^{N} p(x_n)\cos\left(\frac{\pi(m-1)}{2N_{DCT}}(2z_n-1)\right), m = 1, \dots, M \tag{20b}$$

The algorithm for computing the parameters of the regression is as (21). Note that, after convergence, the gradient will be zero which is precisely the constraints.

$$\frac{\partial \mathcal{L}_{ce}}{\partial \lambda_m} = -\sum_{n=1}^{N} \cos\left(\frac{\pi(m-1)}{2N_{DCT}}(2z_n-1)\right)\big(y_n - p(x_n)\big) \tag{21}$$

$$\lambda_m \leftarrow \lambda_m + \mu\left(y_n \cos\left(\frac{\pi(m-1)}{2N_{DCT}}(2z_n-1)\right)\big(y_n - p(x_n)\big)\right)$$

The quality, along with the corresponding polynomial logistic model can be seen in Table 6. Note that the step size for DCT logistic regression is always $0.2/M$, and convergence of the iterative procedure for the coefficients requires less than 400 iterations. In contrast, for the case of polynomial, the step size depends on the energy of the snapshots and becomes very small as $M$ increases. For $M = 5$, the results shown require more than 200.000 iterations. The regression for model orders 2 and 5 are displayed in Figure 5.

**Table 6.** Comparison for polynomial and DCT models in logistic regression.

| $M$ | Model | Iterations | $\text{LL}_{\text{fit}}$ | $R^2$ | $F$ |
|---|---|---|---|---|---|
| 2 | Poly. | 3000 | -6.62 | 0.49 | 8.93 |
| | DCT | 276 | -6.95 | 0.47 | 8.00 |
| 3 | Poly. | 15000 | -6.58 | 0.49 | 4.49 |
| | DCT | 370 | -6.89 | 0.47 | 3.80 |
| 4 | Poly. | $2\cdot10^6$ | -5.82 | 0.55 | 3.34 |
| | DCT | $2\cdot10^3$ | -5.02 | 0.61 | 3.08 |
| 5 | Poly. | $2\cdot10^7$ | -4.89 | 0.62 | 2.82 |
| | DCT | $3\cdot10^3$ | -4.91 | 0.62 | 2.19 |

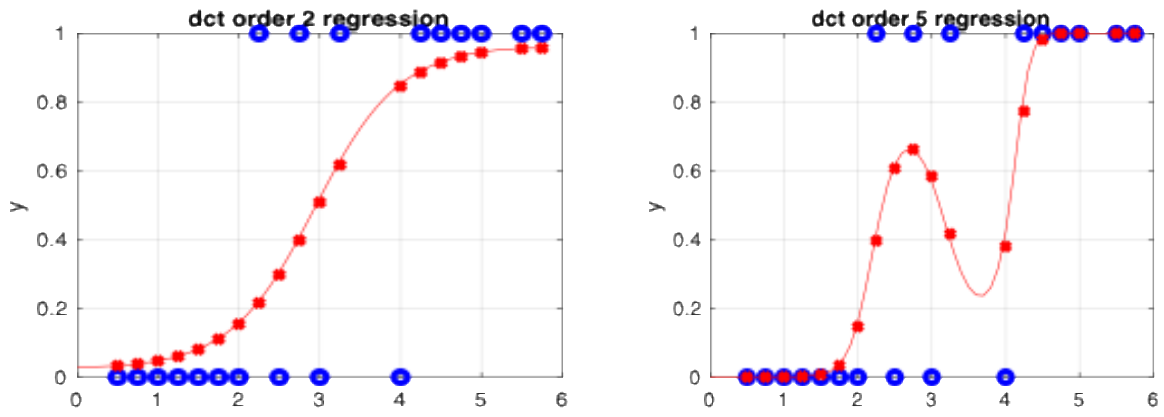

**Figure 5.** DCT Logistic regression. (Left) $M = 2$ and (Right) $M = 5$.

Concerning the impact of the constraints on the regression, the DCT model for logistic regression is more sensitive to outliers than the polynomial counterpart. On the other hand, the iterative computation of the coefficients degrades significantly in the polynomial model with the number of constraints. This is motivated due to the non-orthogonality of the exponential kernels. In contrast, the DCT model benefits from the orthogonality and the bounded nature of the cosine functions, leading to much faster convergence. Furthermore, the DCT does not require fine tuning of the step size.

## V. CONCLUSIONS

This paper has developed a unified variational framework for regression based on the Lagrangian formalism, in which regression problems are formulated as the minimization of an objective function subject to constraints derived from data. Within this formulation, polynomial and logistic regression can all be interpreted as particular instances obtained through different choices of objectives and constraints. This perspective highlights that the resulting model structure is primarily determined by the nature of the constraints.

Traditionally, in the design of learning systems for classification, such as neural networks, the nonlinearity that models the underlying probability distribution is introduced through a logistic or sigmoid activation function. While this choice is well justified by its boundedness and smoothness properties, it is often presented as a heuristic design decision. The variational formulation adopted in this paper provides a formal justification for this choice: when the objective is to maximize the accumulated entropy of the output distribution under moment constraints on the underlying function, the resulting optimal probability distribution is necessarily sigmoid.

The same variational framework enables a broader class of regression models through alternative constraint representations. In particular, the Discrete Cosine Transform (DCT) model arises when the constraints are expressed in the spectral domain rather than in terms of moments. This leads to a regression formulation governed by DCT coefficients of the target function. The resulting representation inherits the boundedness and orthogonality properties of the DCT basis, which improve numerical stability and lead to faster convergence of gradient-based optimization, up to approximately x140 faster in the experiments considered. Moreover, this improved convergence is obtained without the need for critical tuning of the algorithmic parameters, as the step-size remains stable and well behaved across iterations and model orders.

# VI. REFERENCES

[1] D. Montgomery, E. A. Peck, G. Geoffrey Vining. "Introduction to Linear Regression Analysis" (fifth edition). Wiley Series in Probability and Statistics. 2012.

[2] J.P. Hoffman. "Linear regression models". Chapman & Hall/ CRC Statistics in the Social and Behavioral Science 2021.

[3] A. Sen, M. Srivastava. "Regression Analysis". Springer. ISBN 078-0387972114, 1997.

[4] A. I. Perez, M. M. Gost and M. A. Lagunas. "Adaptive function approximation based on the Discrete Cosine Transform (DCT)". CSSC Conference Proceedings, Rodas, Julio 2023.

[5] M.M. Gost, A. I. Perez, M. A. Lagunas. "ENN: A Neural Network with DCT Adaptive Activation Functions". IEEE Journal of Selected Topics in Signal Processing, DOI 101109/JSTSP.2024.3361154. Vol. 18, pp. 232-241, Issue 2, March 2024.

[6] F. Rindler, "Calculus of Variations". Universitext. Springer Cham. ISBN 978-3-319-77636-1. 2018.

[7] Pedro Borges de Melo. "Complex Variational Formulation for Learning Problems". Available at https://Optimization Online, Search for results of the author.

[8] J. Boelaert, E.Ollion. "The Great Regression", Machine Learning, Econometrics, And the Future of Quantitative Social Sciences. Revue française de sociologie, 59(3), 475-506, ISBN 9788272463569. 2018.

[9] Fisher, R. A. (1992), Kotz, Samuel; Johnson, Norman L. (eds.), "Statistical Methods for Research Workers", Breakthroughs in Statistics: Methodology and Distribution, Springer Series in Statistics, New York, NY: Springer, pp. 66–70, doi: 10.1007/978-1-4612-4380-9_6, ISBN 978-1-4612-4380-9.